\documentclass[epj]{svjour}
\usepackage{amssymb,amsmath,graphicx}
\newcommand{\avg}[1]{\langle #1\rangle}
\newcommand{\ol}[1]{\overline{#1}}
\newcommand{\dint}{\int\!\!\!\!\int}
\newcommand{\rmd}{\mathrm{d}}
\newcommand{\req}[1]{(\ref{#1})}

\begin{document}
\title{Heterogeneous network with distance dependent
connectivity}
\author{M. Medo\inst{1,2}
\thanks{matus.medo@unifr.ch}
and J. Smrek\inst{2}
\thanks{jano.smrek@gmail.com}}
\institute{Department of Mathematics,
Physics and Informatics, Mlynsk\'a dolina,
842 48 Bratislava, Slovak republic\and
Department of Physics,
University of Fribourg, 1700 Fribourg, Switzerland}
\abstract{
We investigate a network model based on an infinite regular
square lattice embedded in the Euclidean plane where the node
connection probability is given by the geometrical distance of
nodes. We show that the degree distribution in the basic model
is sharply peaked around its mean value. Since the model was
originally developed to mimic the social network of
acquaintances, to broaden the degree distribution we propose
its generalization. We show that when heterogeneity is
introduced to the model, it is possible to obtain fat tails of
the degree distribution. Meanwhile, the small-world phenomenon
present in the basic model is not affected. To support our
claims, both analytical and numerical results are obtained.
}
\maketitle

\section{Introduction}
Networks are powerful tools for representation of many diverse
systems arising in physics, biology, and sociology. Progress in
this field is rapid; good reviews of our current knowledge are
presented in \cite{Newman,DorMend,BarabAlbert}. In this work we
investigate a network which is embedded in an Euclidean space
where the probability that two nodes are connected by a link
depends on their mutual distance. A similar model was first
proposed by Kleinberg in~\cite{Kleinberg}. Later, a model based
on a regular underlying lattice was proposed and some numerical
results were obtained~\cite{Boguna}; very recently, this work
has been generalized by introducing hidden
variables~\cite{Boguna2}. In~\cite{NisVenk,Petermann}, similar
models with wiring costs depending on distances are studied;
in~\cite{Xulvi,Yook,Manna1,Barth03,Manna2}, the interplay between
geographical distance and node degree is investigated.

Models mentioned above have one feature in common: resulting
networks consist of many short links and long distance
connections are less numerous. Notice that this corresponds to
the picture widely accepted by sociologists investigating
networks of acquaintances~\cite{Gran1,Gran2}. Their key phrase
"Strength of weak ties" has a straightforward interpretation
here: the probability of connecting two nodes must decrease with
the distance slow enough to enable multiple long links. Then
the resulting network resembles the structure observed in the
human society. Our present understanding of this phenomenon
agrees with early mathematical insights presented
in~\cite{Kasturirangan} where importance of multiple edge length
scales was discussed. Notice that also the classical model of
Watts and Strogatz with two types of links~\cite{WattsStrog}
follows a similar pattern.

In this paper we deal with the network model based on the
distance dependent connectivity which was investigated in
\cite{Medo} and is similar to~\cite{Boguna}. This model was
developed to mimic the acquaintance network in a human society.
It allows us to estimate the typical degree of separation
between distant vertices in the network---the results show that
with a proper choice of the dependence between the linking
probability and the nodes distance, the network exhibits the
small-world phenomenon.

However, in the original work the degree distribution $P(k)$ was
not a matter of interest. In this paper we show that it can be
approximated by a Gaussian distribution. This result is not
surprising because the model relies on the edges whose presence
is mutually independent in the same way as it is in the
classical random graph of Erd\"os and R\'enyi. Moreover, we show
that the distribution of $P(k)$ is rather narrow. By contrast,
when we investigate the number of persons's acquaintances in
a~real society, the distribution decays slowly. This observation
and the lack of diversity in the original model were our main
motivations for the presented work. The basic ``homogeneous''
model is generalized by introducing hidden variables which is a
common approach in various network models~\cite{Cald02,Boguna3},
a similar attempt was recently presented in~\cite{Boguna2}. We
investigate the tail behavior of the degree distribution and
show that the resulting network exhibits the small-world
phenomenon.

\section{The basic model}
We assume that nodes of the graph form an infinite square
lattice in the Euclidean plane with the side length of
the elementary square equal to $1$. When modeling a society,
each node represents one person and thus in this way we assume
a~homogeneous distribution of population. The probability that
two vertices with the distance $d$ are connected by an edge we
label as $Q(d)$. Notice that this is the point where we
introduce homogeneity to the network: the probability $Q(d)$ is
the same for every pair of nodes separated by the distance $d$.
The degree $k$ of a node is defined as the number of edges
connected to this node. Consequently, the average node degree
$z:=\avg{k}$ is given by the linking probability $Q(d_i)$
summing over all nodes $i$. When $Q(d)$ changes slowly on the
scale of $1$, the summation can be replaced by an integration
and thus
\begin{equation}
\label{normalization}
z=\sum_i Q(d_i)\approx
\int Q(r)\,\rmd\vec{r}=\int_0^{\infty}Q(r)2\pi r\,\rmd r.
\end{equation}
Notice that with $z$ given, Eq.~\req{normalization} represents a
normalization condition for $Q(d)$.

For a node with the degree $k$, the clustering coefficient $C$
is defined as the ratio $C:=n/\binom{k}{2}$ where $n$ is the
number of edges between the neighbours of the given node. Notice
that $0\leq C\leq1$. For a particular node $X$ and a given
function $Q(d)$, the average value of $n$ is
$\avg{n}=\tfrac12\sum_{i\neq j}Q(d_{Xi})Q(d_{ij})Q(d_{jX})$.
Here the factor $\tfrac12$ corresponds to the fact that by a
plain summation over all $i\neq j$ a doublecounting occurs
($i\leftrightarrow j$). Consequently, the average clustering
coefficient of the network can be approximated as
\begin{equation}
\label{<C>-old}
\avg{C}\approx\frac1{z(z-1)} \dint_{i,j}
Q(d_{Xi})Q(d_{ij})Q(d_{jX})\,\rmd\vec{r}_i\rmd\vec{r}_j.
\end{equation}
Here we again assumed that $Q(d)$ changes slowly on the scale of
$1$.

In~\cite{Medo}, $Q(d)$ was assumed to have the form
\begin{equation}
\label{Q(d)}
Q(d)=\frac1{1+bd^{\alpha}}
\end{equation}
with $\alpha>2$ to allow a proper normalization according to
Eq.~\req{normalization}. This choice was motivated by the
following observations of human society:
\begin{enumerate}
\item When two persons live close to each other, they probably
know each other. Thus we require $Q(0)=1$;
\item The greater is the distance between two persons, the
smaller is the probability that they know each other. Thus
$Q(d)$ should be a decreasing function of $d$;
\item We define the average number of distant people that every
person knows as $N_d\equiv\int_R^{\infty} Q(r)2\pi r\,\rmd r$
where $R$ is large and fixed. We demand $N_d$ sufficiently high
to reflect the observation that many people have distant friends
(e.g.~living on the opposite Earth hemisphere).
\end{enumerate}

For example, $Q(d)=\exp[-bd]$ satisfies (i) and (ii) but if we
choose $R$ that covers half of human population
($\pi R^2\approx 3\cdot 10^9$) and $z\approx200$ (which is a
reasonable value to model real acquaintances), we obtain
$N_d\approx 10^{-11}$ which is effectively zero. Consequently,
Eq.~\req{Q(d)} represents a simple choice of $Q(d)$ which for
$\alpha$ in the range $2.5-3.5$ complies with the requirements
written above. Yet, we do not claim that these three
observations allow us to guess the precise form of $Q(d)$. We
merely suppose that our choice is able to capture basic features
of the human acquaintances network. More detailed discussion on
the nature of $Q(d)$ can be found in \cite{Medo}.

In addition to the clustering coefficient defined above, another
important characteristics of random networks is the degree of
separation (or equally the shortest path length). It is defined
as the minimal number of vertices along the shortest path
between two given nodes. Denoting the geometrical distance of
these two nodes as $l$, in the original paper it was shown that
for $Q(d)$ given by Eq.~\req{Q(d)}, the typical degree of
separation of distant nodes is
\begin{equation}
\label{D-old}
\tilde D(l)\approx -\frac{\ln Q(l)}{\ln z}.
\end{equation}
Since in two dimensions, the typical distance $l$ scales with
the network size $S$ as $\sqrt{S}$. Consequently, for the
distance dependence given by Eq.~\req{Q(d)}, the typical
topological distance of two nodes in the network scales as
$\tilde D\sim\ln S$. For the human acquaintances network is
$S=6\cdot 10^9$ and hence $l\approx80\,000$; when $\alpha$ lies
in the range $2.5-3.5$, and $z$ in the range $50-500$, we obtain
$\tilde D$ in the range $3-10$. In addition, by numerical
integration of Eq.~\req{<C>-old}, for the described parameters
the mean clustering coefficient lies in the range $0.05-0.30$.
We can conclude that the given network exhibits the small-world
phenomenon.

\section{The degree distribution}
Let's choose one node of the network, we label it as $X$. The
plane can be divided into thin concentric rings centered at $X$.
If the ring radius is $r$ and its width is $w$, it covers
approximately $N=2\pi rw$ vertices. Meanwhile, all
vertices in one ring have approximately the same distance from
$X$. Therefore they also have approximately the same probability
$Q(r):=p$ to be connected with $X$. Since links are drawn
independently, the number of neighbours of $X$ in the ring with
radius $r$, $n(r)$, is a random quantity with the binomial
distribution whose mean is $Np$ and the variance is
$Np(1-p)=2\pi rw Q(r)[1-Q(r)]$.

The degree $k$ of node $X$ is obtained by summing $n(r)$ over
all rings. The central limit theorem applies here and thus $k$
is normally distributed and its variance $\sigma_k^2$ is the sum
of variances of all contributions $n(r)$. Replacing the
summation over all rings by the integration we obtain
\begin{equation}
\label{sigma}
\sigma_k^2\approx
\int_0^{\infty}2\pi r Q(r)\big[1-Q(r)\big]\,\rmd r=2z/n.
\end{equation}
To confirm this result numerically, in tab.~\ref{tab-sigma} the
quantity $n\sigma_k^2/z$ is shown for various values of $z$ and
$n$. As can be seen, the numerical results are well approximated
by the analytical prediction $n\sigma_k^2/z=2$ for a wide range
of parameters.
\begin{table}
\centering
\begin{tabular}{r|ccc}
                & $z=50$ & $z=150$ & $z=500$\\
\hline
\vbox to 9pt{}
$\alpha=2.5$    & 1.96   & 1.94    & 1.88\\
$\alpha=3.0$    & 1.97   & 2.03    & 1.93\\
$\alpha=3.5$    & 2.00   & 2.05    & 2.00
\end{tabular}
\caption{Numerical estimates of $n\sigma_k^2/z$ for various
values of $\alpha$ and $z$ on the square lattice with the
dimensions $1\,000\times1\,000$ ($4\,000\times4\,000$ for
$\alpha=2.5$), the variance $\sigma_k^2$ is obtained from
10\,000 realisations of the model.}
\label{tab-sigma}
\end{table}

We can conclude that the node degree $k$ has approximately the
Gaussian distribution with the mean $z$ and the variance $2z/n$.
For values of $z$ resembling a real society ($z$ of the order of
hundreds) it follows that $\sigma_k\ll z$ and thus the degree
distribution is sharply peaked around its mean value (narrowness
of the degree distribution is clearly visible in
Fig.~\ref{fig-Pk}). This is in a clear contradiction with the
empirical studies~\cite{Ebel,cinania,Csanyi,Onnela} which
suggest power-law behavior. The resulting social network is
strongly homogeneous---it lacks nodes exceeding others in degree
by orders of magnitude. In the following section we investigate
how this basic model can be modified to produce a~heterogeneous
network and exhibit a broad degree distribution.

\section{Heterogeneous network model}
The probability distribution $Q(d)$ given by Eq.~\req{Q(d)},
fundamental for this model, has two natural parameters: $b$ and
$\alpha$. Heterogeneity can be introduced to the network by
assigning random values of these parameters to each node (with
the constraints $b>0$, $\alpha>2$). To keep the acquaintance
relation symmetric we symmetrize the probability $Q_{ij}(d)$
that persons $i$ and $j$ with the distance $d$ know each other
by the relation
\begin{equation}
\label{symmetrization}
Q_{ij}(d;b_i,\alpha_i,b_j,\alpha_j):=
\frac{Q(d;b_i,\alpha_i)+Q(d;b_j,\alpha_j)}2.
\end{equation}
To simplify our calculations we assume that $\alpha$ is fixed in
the network and only $b$ is a random quantity drawn from the
distribution $\varrho(b)$. The parameter $b$ we call the node
solitariness (as $b$ grows, the number of acquaintances is
decreasing and their average distance is getting smaller). The
average degree of a~vertex with the solitariness $b$ is now
\begin{eqnarray}
\label{k-mix}
z(b)&=&\!\int_0^{\infty}\frac{Q(r;b)+Q(r;b')}2
\,2\pi r\varrho(b')\,\rmd r\,\rmd b'=\nonumber\\
&=&\!\int_0^{\infty}\!\!\!\!Q(r;b)\pi r\,\rmd r+
\frac{\avg{z(b)}}2=
\frac{\pi^2b^{-2/\alpha}}{\alpha\sin(2\pi/\alpha)}+\frac z2.
\end{eqnarray}
Here we again replaced the summation by an integration;
$\avg{z(b)}$ is the average connectivity in the network which we
already labeled as $z$. As the solitariness $b$ of a vertex goes
to zero, $z(b)$ goes to infinity. By contrast, as $b$ increases
to infinity, $z(b)$ has a lower bound which is equal to $z/2$.

According to our previous discussions, we would like to
generalize the model to exhibit a wide connectivity
distribution. To achieve this, high-degree nodes with small
values of $b$ must be present. However, the value $b=0$ is
pathological for it makes the probability distribution
$Q(d;b)$ flat and creates a node with an infinite degree (if the
network itself is infinite). Thus for the distribution
$\varrho(b)$ we require $\varrho(0)=0$. The simplest possible
choice is $\varrho(b)=Kb^{\beta}$ for $b\in(0;B]$, $\beta>0$.
Values of $K$ and $B$ are fixed by the condition $\avg{z(b)}=z$
and by the normalization of $\varrho(b)$, leading to
\begin{equation}
\label{ro-b}
\begin{aligned}
B&=\Bigg[\frac{(\alpha+\alpha\beta-2)z\sin(2\pi/\alpha)}
{(1+\beta)2\pi^2}\Bigg]^{-\alpha/2},\\
K&=(1+\beta)\,B^{-(1+\beta)}
\end{aligned}
\end{equation}
Since high degrees are due to small values of $b$, the results
derived below hold for all $\varrho(b)$ which can be
approximated by $\varrho(b)\sim b^{\gamma}$ for $b$ small.
Thanks to the constraint $\varrho(0)=0$ and the Taylor expansion,
this is already a quite general class of functions. However, in
this paper we focus on the power-law $\varrho(b)$ which allows
us to investigate the model analytically.

\begin{figure}
\centering
\includegraphics[scale=0.3]{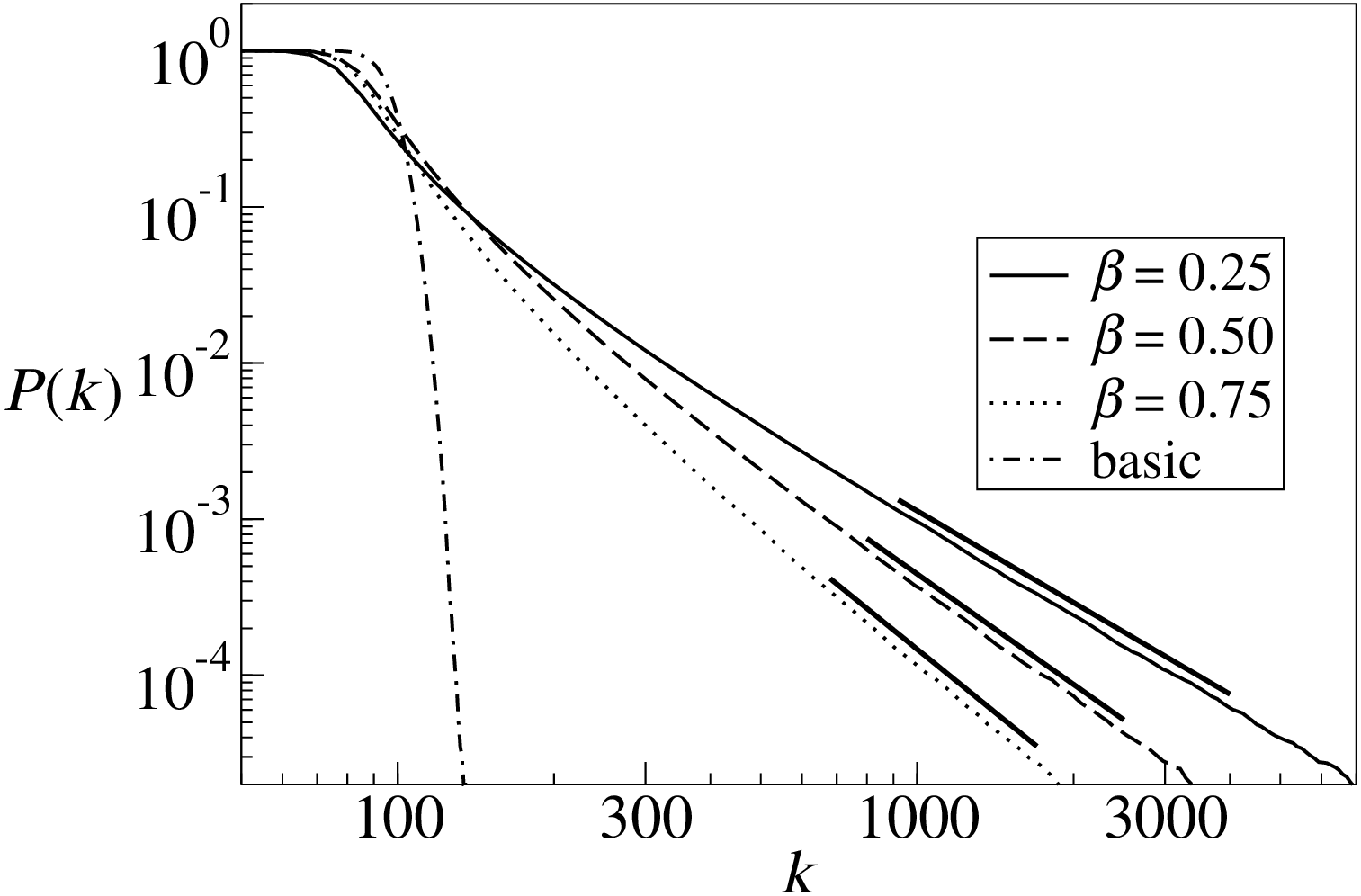}
\caption{The cumulative degree distribution $P(k)$ plotted in
a~logarithmic scale. The thick lines have slopes $1.95$, $2.35$,
and $2.70$ respectively (values predicted by Eq.~\req{power-law}
are $1.88$, $2.25$, and $2.63$ respectively). The probability
distributions were obtained by 20 realisations of the model on
the lattice with dimensions $600\times600$ for $\alpha=3$ and
$z=100$. For a comparison, the degree distribution of the basic
homogeneous model is also shown.}
\label{fig-Pk}
\end{figure}

First we show that the chosen form of $\varrho(b)$ leads to the
desired fat distribution of connectivities. For the distribution
of $z(b):=\ol{k}$ we have
$$
g(\ol{k})=\varrho(b)\Big/\left\vert
\frac{\rmd\ol{k}}{\rmd b}\right\vert=
\frac{2K}{\alpha}\,b^{1+\beta+2/\alpha}.
$$
Consequently, using Eq.~\req{k-mix} and assuming $\ol{k}\gg z$
we obtain
$$
g(\overline{k})\sim\overline{k}^{\,-(1+\alpha/2+\alpha\beta/2)}.
$$
We already know that when $b$ is given, the probability
distribution of the vertex degree is sharply peaked. Therefore
we can approximate the distribution $f(k)$ which we are
searching for by the distribution $g(\ol{k})$ of the mean
degree. Then we obtain
\begin{equation}
\label{power-law}
f(k)\sim k^{-(\alpha\beta+\alpha+2)/2},\quad
P(k)\sim k^{-(\alpha\beta+\alpha)/2}.
\end{equation}
Here $P(k)$ is the cumulative probability distribution of the
vertex degree. We see, that for the chosen $\varrho(b)$, the
degree distribution has a power-law tail. In Fig.~\ref{fig-Pk},
this analytical result is compared with a numerical simulation
of the model for $z=100$ and $\alpha=3$. The power-law character
of $P(k)$ is clearly visible for $k\gtrsim300$ and the
approximate values of the power-law exponents confirm
Eq.~\req{power-law}.

\begin{figure}
\centering
\includegraphics[scale=0.3]{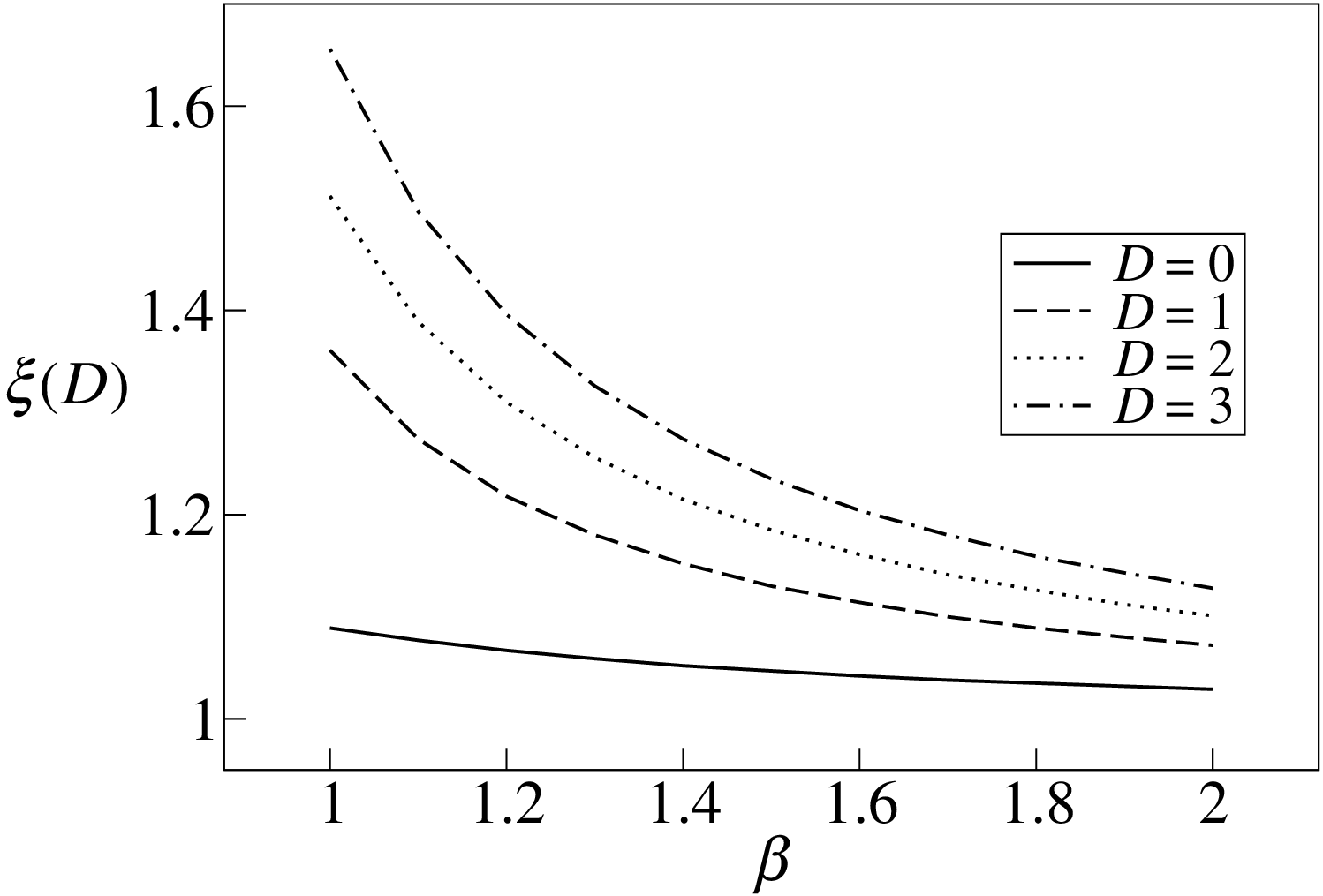}\\[6pt]
\includegraphics[scale=0.3]{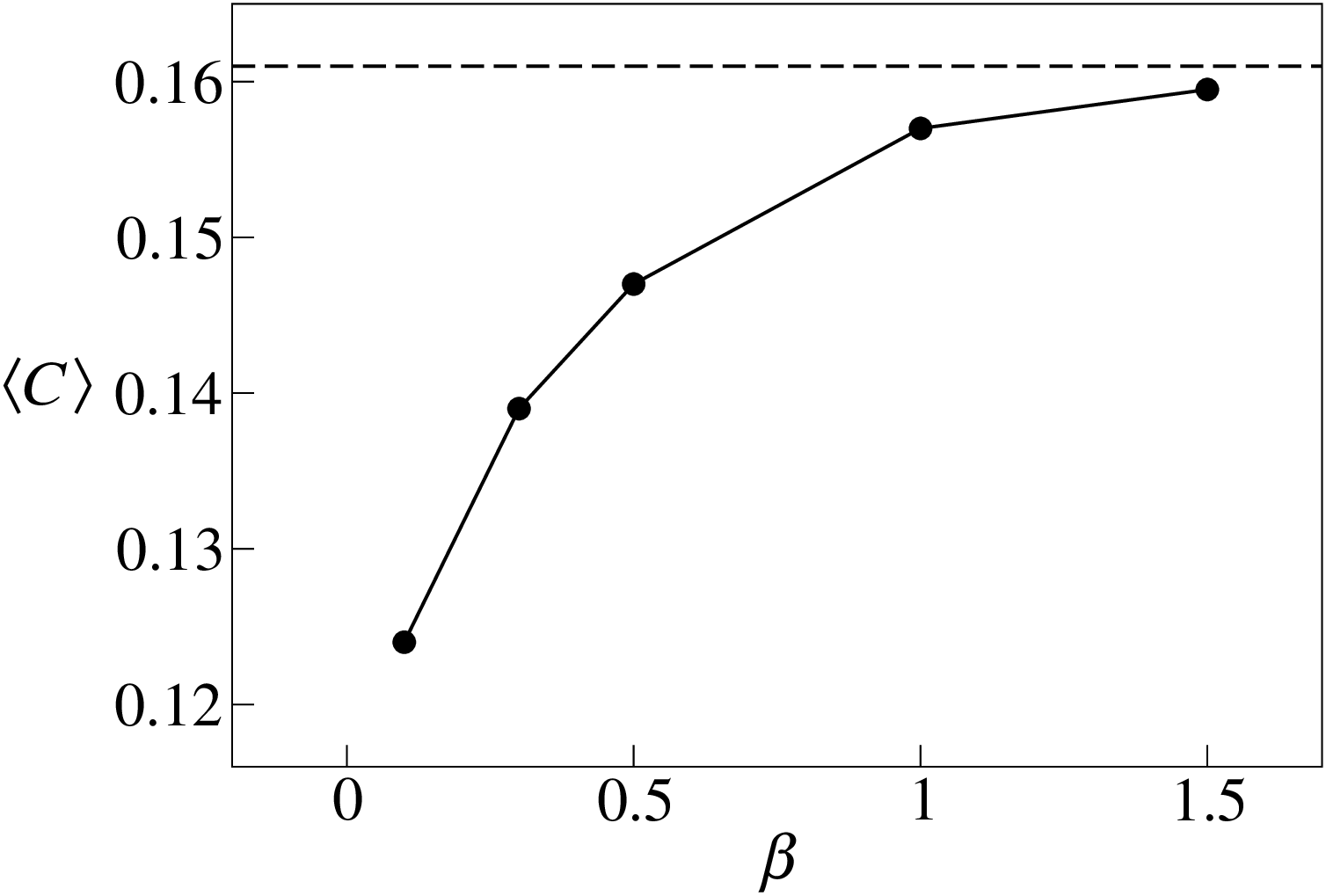}
\caption{Changes of the the main network properties with $\beta$
for $z=300$ and $\alpha=3$. In the upper figure, the ratio
$\xi(D):=P(D)_{\mathrm{HE}}/P(D)_{\mathrm{HO}}$ is drawn
according to Eq.~\req{PD-povodne} and Eq.~\req{PD-nove}.
The course of $\avg{C}$ in the lower figure has been obtained by
a numerical simulation of the model with averaging over 1\,000
realizations; the dashed line presents the limiting value of the
clustering coefficient in the limit $\beta\to\infty$.}
\label{fig-cc-q}
\end{figure}

Now we show that the modified network model still exhibits the
small world phenomenon. The probability that two vertices with a
fixed distance $l$ have the degree of separation $D$ we label as
$P(D)$. We can examine this quantity by techniques similar to
those presented in~\cite{Medo}. There it was shown that in the
resulting homogeneous network, the first approximation of $P(D)$
has the form
\begin{equation}
\label{PD-povodne}
P(D)_{\mathrm{HO}}\approx (D+1)z^D Q(l).
\end{equation}
The derivation of a similar result $P(D)_{\mathrm{HE}}$ for the
heterogeneous network model proposed here can be found in
Appendix A; it is well defined only when $\alpha\beta>2$.
In Fig.~\ref{fig-cc-q}, the resulting ratio
$\xi(D):=P(D)_{\mathrm{HE}}/P(D)_{\mathrm{HO}}$ is shown as
a~function of $\beta$. Notice that in the limit $\beta\to\infty$
all ratios $\xi(D)$ approach to $1$. This is because as $\beta$
increases, a higher weight is given to values of $b$ close to
the upper bound $B$. In particular, in the limit
$\beta\to\infty$ all nodes share the same value of solitariness,
$B$. Thus we can say that the proposed generalization is in the
limit $\beta\to\infty$ equivalent to the original model. One can
notice that $\xi(D)>1$ for all $D$. This means that in the
proposed heterogeneous network the probability to find a path of
a certain length is higher than in the homogeneous network. In
other words, hubs (nodes with a high degree) present in the
heterogeneous network facilitate formation of short paths.
Therefore, for the typical degrees of separation the inequality
$\tilde D_{\mathrm{HE}}<\tilde D_{\mathrm{HO}}$ holds. On the
other hand, since the ratios $P(D+1)/P(D)$ (which are of order
of $z$) are much larger then the ratios $\xi(D)$ shown in
Fig.~\ref{fig-cc-q}, we can also say that the introduction of
heterogeneity to the network does not change the typical degree
of separation substantially and
$\tilde D_{\mathrm{HE}}\approx\tilde D_{\mathrm{HO}}$.

The average clustering coefficient $\avg{C}$ cannot be trea\-ted
analytically and therefore in Fig.~\ref{fig-cc-q} we present
only numerical results. They confirm the expected fact that
$\avg{C}$ is little sensitive to changes of the model
parameters and thus it is almost as high as in the original
model. One can also notice that with increasing $\beta$,
$\avg{C}$ approaches to the value $0.161$ valid for the original
model (this value is taken from \cite{Medo}, in
Fig.~\ref{fig-cc-q} it is shown as a dashed line). This limit
behavior is similar to the limit behavior of $\xi(D)$. Since we
observe both a small typical degree of separation and a high
average clustering coefficient, the heterogeneous network
exhibits the small world phenomenon.

\section{Conclusion}
In this paper we investigated a network model where links are
drawn according to nodes distances. Building on the basic
model~\cite{Medo}, we proposed a~generalization aiming to
introduce heterogeneity to the network and also fat tails to the
degree distribution. First, a hidden random parameter $b$ is
assigned randomly to each vertex of the network. Then between a
pair of nodes, a link is drawn with the probability depending on
the hidden parameter values of these two nodes. As a result we
obtain highly heterogeneous network which exhibits a power-law
distribution $P(k)$ over a large range of connectivities. With
respect to the social interpretation of the model, one can say
that it produces a social network where highly sociable party
goers are present along with loners. The proportion of highly
connected nodes can be adjusted by the distribution, from which
the hidden parameter $b$ is drawn---in this work we focused on
the simple distribution $\varrho(b)=Kb^{\beta}$. We also showed
that for the resulting network, the typical degree of separation
is small and the average clustering coefficient is high; both are
approximately equal to the corresponding values for the
homogeneous network with same $z$ and $\alpha$. Thus we conclude
the small world phenomenon is present in the networks produced
by the proposed model.

\vspace{-12pt}
\begin{acknowledgement}
MM would like to thank the Universit\'e de Fribourg for
the financial support and kind hospitality and to Paulo Laureti
for enjoyable and helpful discussions. JS would like to thank
Mat\'u\v s Medo for introduction to the field and patient and
agreeable collaboration.
\end{acknowledgement}

\appendix
\section{Calculation of $P(D)$ in the heterogeneous network}
To obtain an approximate expression for $P(D)$, the treatment is
similar to the treatment of the homogeneous network model
in~\cite{Medo}. As we will see, differences and complications
arise from the additional averaging over possible values of the
solitariness $b$ with $\varrho(b)$.

\begin{figure}
\centering
\includegraphics[scale=0.8]{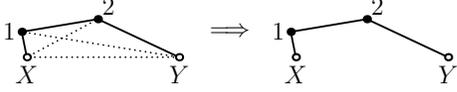}
\caption{The exact diagram for $P(2)$ (left) and the approximate
one utilizing the properties of $Q(d)$ (right).}
\label{fig-diagrams}
\end{figure}

We pick two nodes with a large geometrical distance $l$, let's
label them $X$ and $Y$. As an illustrative example we examine the
probability $P(2)$. That is, we examine paths between $X$ and $Y$
that have two intermediate vertices (Fig.~\ref{fig-diagrams},
left). Notice that $D=2$ requires that links $X2$, $1Y$, and $XY$
are not present. Since probabilities of these links are small, to
obtain a first approximation of $P(D)$ we neglect that such
shortcuts may occur. Consequently, the diagram for $P(2)$ is
simplified (Fig.~\ref{fig-diagrams}, right) to the existence of
edges $X1$, $12$, and $2Y$.

\begin{figure}
\centering
\includegraphics[scale=0.8]{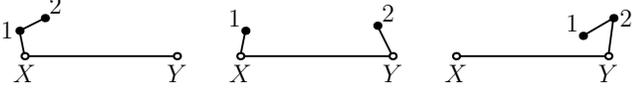}
\caption{Three diagrams contributing substantially to $P(2)$.}
\label{fig-diagrams2}
\end{figure}

Another simplification comes from the form of $Q(d)$ given by
Eq.~\req{Q(d)}. It is easy to check that when $l$ is large, for
$d\ll l$ holds $Q(l-d)Q(d)\gg Q(l/2)Q(l/2)$. This means that
among all paths $X12Y$, the fundamental contribution comes from
those which contain only one long link. Moreover, for $d\ll l$
we have also $Q(l-d)\approx Q(l)$ and therefore the probability
of the long link can be approximated by $Q(l)$. As a result we
can further simplify the right diagram in Fig.~\ref{fig-diagrams}
to Fig.~\ref{fig-diagrams2} where the only three diagrams
substantially contributing to $P(2)$ are shown (three different
possibilities appear because there are three ways to choose the
long link in the path $X12Y$; since probability of the long
link is always approximately equal to $Q(l)$, the link is drawn
between $X$ and $Y$).

In the basic network model, the contribution of the left most
diagram in Fig.~\ref{fig-diagrams2} to $P(2)$ is
$$
P_1=\dint_{1,2} Q(d_{X1})Q(d_{12})Q(l)\,\rmd\vec{r}_{X1}
\rmd\vec{r}_{12}=Q(l)z^2.
$$
Since the remaining two diagrams give the same result, together
we have $P(2)\approx 3z^2Q(l)$ which agrees with
Eq.~\req{PD-povodne}. In the modified model of a heterogeneous
network, $Q(d)$ is generalized to $Q(d;b)$ and the connection
probability is symmetrized by $[Q(d;b_1)+Q(d;b_2)]/2$. Then for
the left most diagram shown in Fig.~\ref{fig-diagrams2} we
encounter the complex expression
$[Q(d_{X1};b_X)+Q(d_{X1};b_1)]\times
[Q(d_{12};b_1)+Q(d_{12};b_2)]\times
[Q(l;b_2)+Q(l;b_Y)]$.
Moreover,  in addition to the integration over
$\vec{r}_{X1},\vec{r}_{12}$, we also have to integrate over
$b_X,b_Y,b_1,b_2$. Then we encounter the following integrals
\begin{eqnarray*}
\label{konstanty-K}
\int_0^B
\ol{k}(b)\varrho(b)\,\rmd b&:=&z,\\
\int_0^B\! Q(l;b)\varrho(b)\,\rmd b\approx
\frac{\beta+1}{\beta}\,Q(l;B)&:=& K_1 Q(l;B)\\
\int_0^B\!\!\!\int_1 Q(r_1;b)Q(l;b)\varrho(b)
\,\rmd b\,\rmd\vec{r}_1\approx\\
\approx\frac{\alpha+\alpha\beta-2}{\alpha\beta-2}\,z
Q(l;B)&:=&
K_2 z Q(l;B),\\
\int_0^B\!\!\!\dint_{1,2}
Q(r_1;b)Q(r_2;b)\varrho(b)\,\rmd b\,\rmd\vec{r}_1\,
\rmd\vec{r}_2\approx\\
\approx\frac{(\alpha+\alpha\beta-2)^2}
{\alpha(1+\beta)(\alpha+\alpha\beta-4)}\,z^2&:=& K_3 z^2.
\end{eqnarray*}
The third integral converges when $\alpha\beta>2$, the fourth
when $\alpha+\alpha\beta>4$ (since $\alpha>2$, this is a weaker
restriction).

Using the steps and notation introduced above we finally obtain
the approximate result
\begin{equation}
\label{PD-nove}
P(D)_{\mathrm{HE}}\approx\frac{L(D)}{2^D}\,Q(l;B)z^D
\end{equation}
where
\begin{eqnarray*}
L(0)&=&K_1,\\
L(1)&=&3K_1+K_2,\\
L(2)&=&6K_1+4K_2+2K_1K_3,\\
L(3)&=&10K_1+10K_2+10K_1K_3+2K_2K_3,\\
L(4)&=&15K_1+20K_2+30K_1K_3+12K_2K_3+3K_1K_3^2,\dots
\end{eqnarray*}
For $D>4$ we obtain even more complicated expressions.
Nevertheless, since the typical degree of separation is small in
the discussed model, this is not a crucial complication and the
solution is tractable. We can notice that in the limit
$\beta\to\infty$ we have $K_1,K_2,K_3\to1$ and therefore
$P(D)_{\mathrm{HE}}=P(D)_{\mathrm{HO}}$ as expected.


\begin{thebibliography}{99}
\bibitem{Newman} M.E.J. Newman,
SIAM Rev. {\bf 45}, 167 (2003)

\bibitem{DorMend} S.N. Dorogovtsev and J.F.F. Mendes,
Adv. Phys. {\bf 51}, 1079 (2002)

\bibitem{BarabAlbert} R. Albert and A.-L. Barab\'asi,
Rev. Mod. Phys. {\bf 74}, 47 (2002)

\bibitem{Kleinberg} J. Kleinberg,
in {\it Proceedings of the 32nd ACM Symposium on Theory of
Computing}, edited by F. Yao and E. Luks (ACM, New York, 2000),
p. 163

\bibitem{Boguna} M. Bogu\~n\'a et al,
Phys. Rev. E {\bf 70}, 056122 (2004)

\bibitem{Boguna2} M. \'Angeles Serrano et al,
Phys. Rev. Lett. {\bf 100}, 078701 (2008)

\bibitem{NisVenk} M. Nisha and G. Venkatesh,
Phys. Rev. E {\bf 63}, 021117 (2001)

\bibitem{Petermann} T. Petermann and P. De Los Rios,
Phys. Rev. E {\bf 73}, 026114 (2006)

\bibitem{Xulvi} R. Xulvi-Brunet and I.M. Sokolov,
Phys. Rev. E {\bf 66}, 026118 (2002)

\bibitem{Yook} S.-H. Yook et al,
PNAS {\bf 99}, 13382 (2002)

\bibitem{Manna1} S.S. Manna and S. Parongama,
Phys. Rev. E {\bf 66}, 066114 (2002)

\bibitem{Barth03} M. Barth\'elemy,
Europhys. Lett. {\bf 63}, 915 (2003)

\bibitem{Manna2} G. Mukherjee and S.S. Manna,
Phys. Rev. E {\bf 74}, 036111 (2006)

\bibitem{Gran1} M.S. Granovetter,
Am. J. of Sociology {\bf 78}, 1360 (1973)

\bibitem{Gran2} M.S. Granovetter,
Sociological Theory {\bf 1}, 201 (1983)

\bibitem{Kasturirangan} R. Kasturirangan,
{\it Preprint} cond-mat/9904055 (1999)

\bibitem{WattsStrog} D.J. Watts and S. Strogatz,
Nature {\bf 393}, 440 (1998)

\bibitem{Medo} M. Medo,
Physica A {\bf 360}, 617 (2006)

\bibitem{Cald02} G. Caldarelli et al,
Phys. Rev. Lett. {\bf 89}, 258702 (2002)

\bibitem{Boguna3} M. Bogu\~n\'a and R. Pastor-Satorras,
Phys. Rev. E {\bf 68}, 036112 2003

\bibitem{Ebel} H. Ebel, L.-I. Mielsch and S. Bornholdt,
Phys. Rev. E {\bf 66}, 035103 (2002)

\bibitem{cinania} T. Zhou et al,
Phys. Rev. E {\bf 76}, 037102 (2007)

\bibitem{Csanyi} G. Cs\'anyi and B. Szendr\"oi,
Phys. Rev. E {\bf 69}, 036131 (2004)

\bibitem{Onnela} J.P. Onnela et al,
New J. Phys. {\bf 9}, 179 (2007)
\end{thebibliography}
\end{document}